\documentclass[prd,twocolumn,showpacs,amsmath,amssymb,floatfix]{revtex4}
\usepackage{graphicx,color,dcolumn,booktabs}
\usepackage{longtable,lscape}
\usepackage{txfonts}
\usepackage{amssymb}
\usepackage{indentfirst}

\begin{document}
\title{Newly observed $D(2550)$, $D(2610)$, and $D(2760)$ as $2S$ and $1D$ charmed mesons}

\author{Zhi-Feng Sun$^{1,2}$}\author{Jie-Sheng Yu$^{1,2}$}
\author{Xiang Liu$^{1,2}$
}
\email{xiangliu@lzu.edu.cn}

\affiliation{$^1$School of Physical Science and Technology, Lanzhou University, Lanzhou 730000,  China\\
$^2$Research Center for Hadron and CSR Physics,
Lanzhou University $\&$ Institute of Modern Physics of CAS, Lanzhou 730000, China}

\author{Takayuki Matsuki}
\email{matsuki@tokyo-kasei.ac.jp}
\affiliation{Tokyo Kasei University, 1-18-1 Kaga, Itabashi, Tokyo 173-8602, Japan}

\date{\today}
\begin{abstract}

We study three newly observed $D(2550)$, $D(2610)$, and $D(2760)$ by the BaBar Collaboration utilizing the mass spectra and investigating the strong decays. Our calculation indicates that $D(2610)$ is an admixture of $2^3S_1$ and $1^3D_1$ with $J^P=1^-$. $D(2760)$ can be explained as either the orthogonal partner of $D(2610)$ or $1^3D_3$.
{Our estimate of the decay width for $D(2550)$, assuming it as $2^1S_0$, is far below the experimental value.}
The decay behavior of the remaining two $1D$ charmed mesons, i.e., $^3D_2$ and $^1D_2$ ($J^P=2^-$) states, is predicted, which will help future experimental search for these missing $D$-wave charmed mesons.

\end{abstract}

\maketitle

Very recently the BaBar Collaboration has observed new charmed mesons, $D(2550)$, $D(2610)$, and $D(2760)$ \cite{BABAR}. The spectra of $D$ mesons are very poorly known because higher states have been hindered by poor statistics and their relatively large widths.
The BaBar analysis on these particles is that $2S$ and $1D$ are the most likely candidates 
when relying on the mass values predicted by
the conventional nonrelativistic potential model \cite{GI} and the relativistic potential models \cite{MMS,PE,EFG}.
By reanalyzing their data, especially by studying their decay widths with successful $^3P_0$ model \cite{Micu:1968mk}, we would like to see whether the quark model prediction fits with their data.

In the $D^+\pi^-$ invariant mass spectrum, two $D$ mesons, $D(2610)^0$ and $D(2760)^0$ with neutral charge,
have been observed along with confirming two established charmed mesons, $D_0^*(2400)^0$ and $D_2^*(2460)^0$. BaBar has also found the isospin partners $D(2610)^+$ and $D(2760)^+$ in $D^0\pi^+$ channel. By analyzing the $D^{*+}\pi^-$ invariant mass spectrum, three structures around $2533.0$ MeV, $2619.0$ MeV,
and $2747.7$ MeV have been released, which shows that BaBar has not only confirmed $D(2610)^0$ and $D(2760)^0$ in $D^{*+}\pi^-$ channel but also has found a new charmed state $D(2550)^0$.
The resonance parameters (in units of MeV) of $D(2550)$, $D(2610)$, and $D(2760)$ are summarized as \cite{BABAR}
\begin{eqnarray}
M_{D(2550)^0}/\Gamma_{D(2550)^0}&=&2539.4\pm4.5\pm6.8/130\pm12\pm13,\nonumber\\
M_{D(2610)^0}/\Gamma_{D(2610)^0}&=&2608.7\pm2.4\pm2.5/93\pm6\pm13,\nonumber\\
M_{D(2760)^0}/\Gamma_{D(2760)^0}&=&2763.3\pm2.3\pm2.3/60.9\pm5.1\pm3.6,\nonumber\label{resonance}\\
M_{D(2610)^+}/\Gamma_{D(2610)^+}&=&2621.3\pm3.7\pm4.2/93,\nonumber\\
M_{D(2760)^+}/\Gamma_{D(2760)^+}&=&2769.7\pm 3.8\pm1.5/60.9,\nonumber
\end{eqnarray}
where the first error is statistical and the second is systematic.

\begin{center}
\begin{figure}[htb]
\begin{tabular}{cccc}
\scalebox{0.375}{\includegraphics{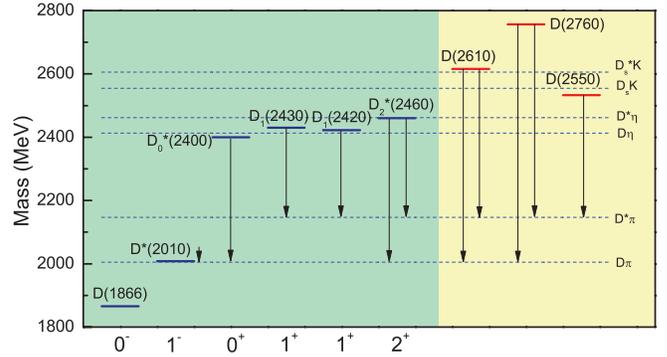}}
\end{tabular}
\caption{(Color online.) The figure shows mass spectrum of six established charmed mesons in PDG \cite{PDG} and three newly observed charmed mesons by the BaBar Collaboration \cite{BABAR} compared with the thresholds of $D^{(*)}\pi$, $D^{(*)}\eta$, $D_0\pi$, and $D_s^{(*)}K$. Here, the experimental decay channels
corresponding to these charmed mesons are also drawn as down arrows. The decay channel $D(2550)\to D_0^*(2400)+\pi$ is omitted
from the figure.\label{mass} }
\end{figure}
\end{center}

As listed in the paper by
the particle data group (PDG) \cite{PDG}, there are six charmed mesons $D$, $D^*$, $D_0^*(2400)$, $D_1(2430)$, $D_1(2420)$, and $D_2^*(2460)$ with the established spin-parity assignments (see Fig. \ref{mass} for details). The newly observed charmed resonances, $D(2550)$, $D(2610)$, and $D(2760)$,
are not only making the spectroscopy of charmed mesons abundant, but also stimulating our interest in revealing the underlying properties of these particles, which, of course, provides a good opportunity to test the existent theory of heavy-light meson system and further enlarges our knowledge of non-perturbative QCD.

The BaBar analysis of angular distribution indicates that $D(2550)$ and $D(2610)$ may be identified as the first radial excitations of $S$-wave charmed mesons while $D(2760)$ can be a $D$-wave state \cite{BABAR}. The quantum number $J^P$ of $D(2550)$ and $D(2610)$ are assigned as $0^-$ and $1^-$, respectively, which explains why BaBar has found $D(2550)$
only in $D^*\pi$ channel and $D(2610)$ both in $D\pi$ and $D^*\pi$ channels. As a candidate of the $D$-wave charmed meson, the spin-parity content of $D(2760)$ can be either $1^-$ or $3^-$ because the observed decay process $D(2760)\to DK$ fully excludes $J^P=2^-$.

{\it Spectroscopy} :
As shown in Table \ref{table1}, different theoretical groups \cite{GI,MMS,PE,EFG} have carried out the systematic calculation of the mass spectra for charmed mesons. In addition to successful reproduction of $1S$ and $1P$ charmed mesons, they predict the masses of charmed mesons with quantum numbers $0^-(2^1S_0)$, $1^-(2^3S_1)$, $1^-(1^3D_1)$, and $3^-(1^3D_3)$. It is noted that the predicted theoretical mass of $D(2550)$ falls into the range $2.483 - 2.589$ GeV, which supports the $0^-(2^1S_0)$ assignment to $D(2550)$.
Due to the fact that the predicted as well as observed masses of charmed mesons for $1^-(2^3S_1)$ and $1^-(1^3D_1)$ are close to each other \cite{BABAR,GI,MMS,PE,EFG}, $D(2610)$ and $D(2760)$ can be regarded as an admixture of $1^-(2^3S_1)$ and $1^-(1^3D_1)$, which satisfies
\begin{eqnarray}\label{mixing1}
 \left ( \begin{array}{ccc} |D(2610)\rangle \\  |D(2760)\rangle\end{array} \right )=
 \left ( \begin{array}{ccc}
  \rm \cos\phi & \rm -\sin\phi \\
  \rm \sin\phi & \rm \cos\phi \end{array} \right )
 \left ( \begin{array}{ccc} |1^-(2^3S_1)\rangle  \\  |1^-(1^3D_1)\rangle  \end{array}\right
),\label{mi}
\end{eqnarray}
where $|1^-(2^3S_1)\rangle$ and $|1^-(1^3D_1)\rangle$ are pure states and $\phi$ denotes the mixing angle.
This situation is similar to that of charmonium $\psi(3770)$, which may be the mixing state of $1^-(2^3S_1)$ and $1^-(1^3D_1)$ charmonia as suggested in Ref. \cite{Rosner:2001nm}.
\renewcommand{\arraystretch}{1.2}
\begin{table}[htb]
\caption{Theoretical calculations for $D$ mesons with
quantum number $n~^{2s+1}L_J$ and a comparison with experimental data. Here, we tentatively set the masses of $1^-(2^3S_1)$ and $1^-(1^3D_1)$ as those of $D(2610)$ and $D(2760)$, respectively. States with double quotations are given by
$|``^3P_1"\rangle=\sqrt{2/3}|^3P_1\rangle-\sqrt{1/3}|^1P_1\rangle$ and
$|``^1P_1"\rangle=\sqrt{1/3}|^3P_1\rangle+\sqrt{2/3}|^1P_1\rangle$ \cite{MMS}, which correspond to two $1^+$ states in $S$ and $T$ doublets shown in the caption of Table \ref{decay}, respectively. The mass value 2318 MeV for Expt. is for neutral and 2403 MeV is for charged.}
\label{table1}
\begin{center}
\begin{tabular}{cccccccccc}\toprule[1pt]
$J^P\left(n~^{2s+1}L_J\right)$ & Expt.\cite{BABAR,PDG} & GI\cite{GI} &MMS\cite{MMS} &
PE\cite{PE} & EFG\cite{EFG} \\ \midrule[1pt]
$0^-(1^1S_0)$      &1867&1880&1869&1868&1871\\
$1^-(1^3S_1)$      &2008&2004&2011&2005&2010\\
$0^+(1^3P_0)$      &{$\left\{ {\begin{array}{*{20}{c}}
   {2318}  \\
   {2403}  \\
   \end{array}} \right.$}
                       &2400&2283&2377&2406\\
$1^+(1``{^3P_1}")$&2427&2490&2421&2417&2469\\
$1^+(1``{^1P_1}")$&2420&2440&2425&2460&2426\\
$2^+(1^3P_2)$      &2460&2500&2468&2490&2460\\
$0^-(2^1S_0)$      &2533&2580&2483&2589&2581\\
$1^-(2^3S_1)$      &2619&2640&2671&2692&2632\\
$1^-(1^3D_1)$      &2763&2820&2762&2795&2788\\
$3^-(1^3D_3)$      & ? & 2830 & - &2799&2863\\
\bottomrule[1pt]
\end{tabular}
\end{center}
\end{table}

From the above analysis and Table \ref{table1}, in general one concludes that $2^1S_0$ is a good candidate for pseudoscalar meson $D(2550)$. $D(2610)$ may be a pure $2^3S_1$ state or an admixture of $2^3S_1$ and $1^3D_1$ states. $D(2760)$ can be a pure $1^3D_1$ or a pure $1^3D_3$ or an admixture of $2^3S_1$ and $1^3D_1$ as the the orthogonal cousin of $D(2610)$.

{\it Decay width} :
Further studies of the two-body strong decay will be helpful to distinguish the different structure assignment to $D(2550)$, $D(2610)$, and $D(2760)$ using the quark pair creation (QPC) model \cite{Micu:1968mk,LeYaouanc:1977gm,LeYaouanc:1988fx}, which is a successful phenomenological model to deal with the Okubo-Zweig-Iizuka (OZI) allowed strong decays of hadron. The relevant decay channels of these particles are listed in Table \ref{decay} applying the quantum number assignment to $D(2550)$, $D(2610)$, and $D(2760)$ discussed above.
Defining the transition operator $\mathcal{T}$ in the QPC model \cite{Micu:1968mk,LeYaouanc:1977gm,LeYaouanc:1988fx}, the main task is to calculate the helicity amplitude $M^{M_{J_A}M_{J_B}M_{J_C}}(\mathbf{K})$
corresponding to the strong decay processes $A(c(1)\bar{q}(2))\to B(c(1)\bar{q}(3))+C(\bar{q}(2)q(4))$ shown in Table \ref{decay}, where the harmonic oscillator (HO) wave function $\Psi_{n_r\ell m}(\mathbf{k}) = \mathcal{R}_{n_r\ell}(R,\mathbf{k})\mathcal{Y}_{n_r\ell m}(\mathbf{k})$ is applied to calculate the spatial integral in the transition matrix element (See Ref. \cite{Luo:2009wu} for more details).
The parameter $R$ in the HO wave function is adjusted so that it reproduces the realistic root mean square (RMS) radius.
The RMS is obtained by solving the Schr\"{o}dinger equation with the
potential in Refs. \cite{Close:2005se,Li:2009qu}, which gives different $R$ values corresponding to $\pi/\eta$, $\rho/\omega$, $K$, $D$, $D^*$, $D_s$, $D_s^*$, $D_1(2430)$, $D_1(2420)$, $D_2^*(2460)$, $D(2^3S_J)$, and $D(1^3D_J)$, respectively. The remaining input parameters are the constituent quark masses of charm, up/down, and strange, i.e.,  1.45 GeV, 0.33 GeV, and 0.55 GeV, respectively \cite{Li:2009qu}. In addition, the strength of the QPC from the vacuum can be extracted by fitting the data. In this letter, we take $\gamma=6.3$ \cite{Godfrey:1986wj}. The strength of $s\bar{s}$ creation satisfies
$\gamma_{s}=\gamma/\sqrt{3}$ \cite{LeYaouanc:1977gm}.

\renewcommand{\arraystretch}{1.2}
\begin{table}[htb]
\caption{(Color online.) The allowed decay channels ($\blacksquare$) of $2S$ and $1D$ charmed mesons with the quantum number assignment to $D(2550)$, $D(2610)$, and $D(2760)$ discussed in this letter. For $D(2610)$, decays into $D\rho$, $D\omega$, $D^*\eta$, and $D_s^*K$ are forbidden due to the limit of phase space. Here, $D_1(2430)$ and $D_1(2420)$ are the $1^+$ states in the $S=(0^+,1^+)$ and $T=(1^+,2^+)$ doublets, respectively. \label{decay} }
\begin{center}
\begin{tabular}{ccccccccccccccccc}\toprule[1pt]
      Modes     & Channel             & $0^-(2^1S_0)$         &$1^-(2^3S_1)$      &$1^-(1^3D_1)/3^-(1^3D_3)$\\\midrule[0.7pt]
     $0^-+0^-$  &$D\pi     $    &              &$\blacksquare$&$\blacksquare$ \\
                &$D\eta     $   &             &$\blacksquare$              &$\blacksquare$ \\
                &$D_sK$         &              &$\blacksquare$              &$\blacksquare$ \\
                       $0^-+1^-$  &$D\rho     $ &              &&$\blacksquare$\\
                                  &$D\omega     $   &                          &&$\blacksquare$\\
      $1^-+0^-$  &$D^*\pi$     &  $\blacksquare$            &     $\blacksquare$         &$\blacksquare$\\
                                  &$D^*\eta     $ &              &$\blacksquare$&$\blacksquare$\\
                                  &$D^*_s K$       & & &$\blacksquare$\\
                      $0^++0^-$&$D_0(2400)\pi$   &$\blacksquare$&&\\
                      $1^+(S)+0^-$&$D_1(2430)\pi$       &              &$\blacksquare$&$\blacksquare$\\
                      $1^+(T)+0^-$&$D_1(2420)\pi$       &              &$\blacksquare$&$\blacksquare$\\
                         $2^++0^-$&$D_2^*(2460)\pi$     &              &$\blacksquare$&$\blacksquare$\\
\bottomrule[1pt]
\end{tabular}
\end{center}
\end{table}

In Figs. \ref{NR-1} and \ref{NR-2}, the decay behaviors of $D(2550)$, $D(2610)$, and $D(2760)$ with different quantum number assignments are presented.

{$D(2550)$ :}
The total decay width of $D(2550)$ mainly comes from its $D_0^*\pi^0$ and $D^*\pi$ contributions just shown in Fig. \ref{NR-1}, where the theoretical estimate of the width for $D(2^1S_0)$ is given by $\Gamma\sim 8$ MeV for the typical value $R=3.6$ GeV$^{-1}$, which is {\it far below the observed value of the decay width 127.6 MeV} for $D(2550)$. In Fig. \ref{NR-1}, we also show the $R$ dependence of the total decay width for $D(2^1S_0)$, whose shape is resulted from the node effects of the higher radial wave function. In the range $3.4\le R\le 3.8$ GeV$^{-1}$, the upper limit of the theoretical width is still smaller than the experimental one for $D(2550)$.
{This discrepancy between the theoretical and experimental results may indicate that the quark model calculation might not be appropriate in this case, or the assignment of $2^1S_0$ to $D(2550)$ might be inappropriate. The comparison of the resonance parameters for $D(2550)$ and $D(2610)$ in Eq. (\ref{resonance}) may also be controversial because the width 127.6 MeV for $D(2550)$ is larger than 93 MeV for $D(2610)$ even though the number of decay channels for $D(2610)$ is larger than that of $D(2550)$ (see Table \ref{decay}), where both $D(2550)$ and $D(2610)$ with $J^P=0^-$ and $1^-$, respectively, can decay into $D^*\pi$ through the  $P$-wave. Thus, if assuming that $D(2550)$ as $0^-(2^1S_0)$ charmed meson
and that our method is appropriate, our theoretical estimate for $D(2550)$ becomes a narrow state.

Close and Swanson \cite{Close:2005se} have also studied the strong decay behavior of $2^1 S_0$ charmed meson and have obtained the larger total strong decay width with a smaller discrepancy with experiment. However, the mass of $2^1 S_0$
charmed meson in Ref. \cite{Close:2005se} is taken as $2.58$ GeV (compare this with our value taken from Table \ref{table1}) whose value resulted in the decay width in Ref. \cite{Close:2005se} larger than ours. In our calculation, we follow the approach given by Ref. \cite{Blundell:1996as}, which has been already tested by calculating the decay width for the process $D_{s2}(2573)\to DK+D^*K+D_s\eta$ \cite{Li:2009qu} that is consistent with the corresponding experimental data \cite{PDG}.


\begin{center}
\begin{figure}[htb]
\begin{tabular}{cccc}
\scalebox{0.55}{\includegraphics{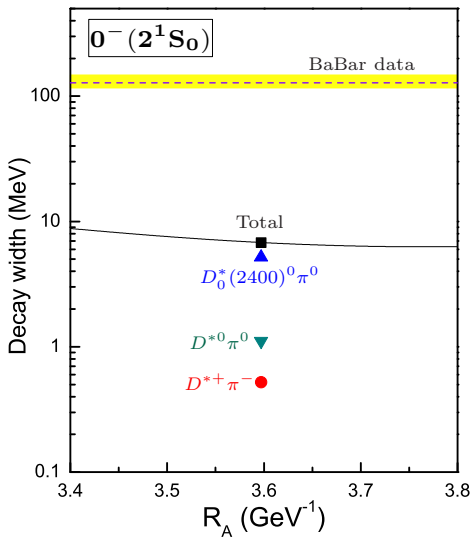}}
\end{tabular}
\caption{(Color online.) The $R$ value dependence and the typical values of the decay width for $D(2^1S_0)$. Here, dashed line with
error band is the BaBar's results of the width of $D(2550)$.
 \label{NR-1} }
\end{figure}
\end{center}


{\it $D(2610)$ and $D(2760)$ }:
As shown in Fig. \ref{NR-2}, assignment of $D(2610)$ and $D(2760)$ to pure $2^3S_1$ and $1^3D_1$ charmed mesons, respectively, can be fully excluded because their decay widths estimated by the QPC model cannot be fitted with the corresponding observed ones when setting the mixing angle $\phi=0$. Instead, there exits unique, possible assignment to the structure of $D(2610)$, i.e., an admixture of $2^3S_1$ and $1^3D_1$ charmed mesons just discussed in Eq. (\ref{mi}). One finds an overlap region (green band) between the theoretical and experimental results for $D(2610)$ with the mixing angle $0.9\le \phi\le 1.5$ radians, which strongly supports that $D(2610)$ and $D(2760)$ are the orthogonal cousins. This explains why $D(2610)$ has been first observed in both $D\pi$ and $D^*\pi$ channels because the main decay modes of of $D(2610)$ are $D_1(2430)\pi$, $D\pi$, $D^*\pi$, and $D\eta$ as shown in Fig. \ref{NR-2}. In addition, theoretical estimate for the several ratios
$\mathcal{R}_1=\Gamma(D^*\pi)/\Gamma(D\pi)$, $\mathcal{R}_2=\Gamma(D\pi)/\Gamma(D_1(2430)\pi)$, and $\mathcal{R}_3=\Gamma(D^*\eta)/\Gamma(D\eta)$ for the dominant decays of $D(2610)$ is
calculated in this mixing angle region $0.9\le \phi\le 1.5$
\begin{eqnarray*}
0.35\le\mathcal{R}_1\le0.47,\; 0.78\le\mathcal{R}_2\le0.91,\; 0.17\le\mathcal{R}_3\le0.33,
\end{eqnarray*}
{which may provide to check the validity of the assumption that the structure assignment to $D(2610)$ is given by Eq. (\ref{mi}). Measurement of the ratio $\Gamma(D^*\pi)/\Gamma(D\pi)$ for $D(2610)$ may be easiest to do because the final states $D^{(*)}$ and $\pi$ can be easily detected.}

{\it $D(2760)$ and $D$-waves }:
If explaining $D(2760)$ as $1^3D_3$, the calculated total decay width for $D(1^3D_3)$ is about a half of the observed one when scanning the range $3.4\le R\le 3.8$ GeV$^{-1}$, where the total widths are $33.2,\,32.9,\,32.5,\,31.9$, and $31.3$ MeV corresponding to $R=3.4,\,3.5,\,3.6,\,3.7$, and $3.8$ GeV$^{-1}$, respectively, which indicates the total decay width is not sensitive to $R$. Owing to the uncertainties coming from the QPC model and the present experiment, $1^3D_3$ assignment to $D(2760)$ cannot be excluded. We especially notice that $D^*\pi$ and $D\pi$ are the dominant decay modes for $D(1^3D_3)$ with the ratio $\Gamma(D^*\pi)/\Gamma(D\pi)=1.1$ with the typical value $R=3.6$ GeV$^{-1}$ (see Table \ref{3-decay}), which explains why $D^{(*)}\pi$ is first observed among many decay channels of $D(2760)$. In addition to this, $D(2760)$ can be the orthogonal partner of $D(2610)$ because the total decay width for $D(2760)$ calculated by the QPC model is close to the upper limit of the observed one with the same mixing angle range as $D(2610)$. This is shown in Fig. \ref{NR-2} with the predicted partial decay behaviors of $D(2760)$. One needs to have further precise measurements on the main decay channels of $D(2760)$, especially on $D^{(*)}\pi$ and $D_{1}(2420)$ because the two-body strong decays of $D(2760)$ for the above two different structure assignments display different behaviors as shown in Table \ref{3-decay} and illustrated in the right diagram of Fig. \ref{NR-2}.

\begin{center}
\begin{figure}[htb]
\begin{tabular}{cccc}
\scalebox{0.75}{\includegraphics{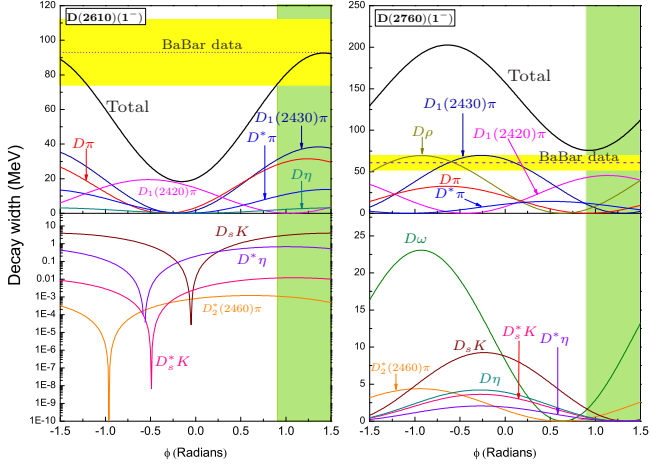}}
\end{tabular}
\caption{(Color online.) The variation of the total and partial decay widths for two $1^-$ states discussed in this letter with the mixing angle range $-1.5\le\phi\le 1.5$ radians. Here, dashed lines with error bands are the BaBar's result of the widths for $D(2610)$ and $D(2760)$.
 \label{NR-2} }
\end{figure}
\end{center}

\renewcommand{\arraystretch}{1.2}
\begin{table}[htb]
\caption{The partial decay width of the $3^-(1^3D_3)$ state (in units of MeV). \label{3-decay} }
\begin{center}
\begin{tabular}{cccc}\toprule[1pt]
 Decay channel  & Decay width&Decay channel  & Decay width\\\midrule[1pt]
 Total           & 32.47 &&\\
 $D^*\pi$        & 12.78 &
 $D\pi$          & 11.71 \\
 $D\rho$         & 3.49 &
 $D_2^*(2460)\pi$& 1.21 \\
 $D\omega$       & 1.07 &
 $D_1(2420)\pi$  & 0.65 \\
 $D\eta$         & 0.55 &
 $D_sK$          & 0.43 \\
 $D^*\eta$       & 0.30 &
 $D_s^*K$        & 0.14 \\
 $D_1(2430)\pi$  & 0.12 \\
 \bottomrule[1pt]
\end{tabular}
\end{center}
\end{table}

{\it $D(2^-)$ states }:
In the following, the decay behaviors of two $2^-$ states in  $1D$ charmed mesons, which are still missing in experiment, are predicted. As described in Fig. \ref{NR-3}, the total and partial decay widths for two $2^-$ states are given with the mass dependence because the spectra of two $2^-$ states are unknown. By increasing the mass of $2^-$ charmed meson, more decay channels are open. As a result, the widths of two $2^-$ charmed mesons become broad, where $\underline{1^-}0^-$ and $\underline{2^+}1^-$ channels (with the underline to mark the quantum number of charmed meson) are dominant decay modes of both $2^-$ charmed mesons for the whole mass region. Additionally, $\underline{0^-}1^-$ is also the dominant decay of $2^-(1``^3D_2")$ charmed meson once this channel is open.
In Fig. \ref{NR-3}, we have not included the decay mode $\underline{0^+}0^-$ because the process $2^-\to \underline{0^+}0^-$ through $D$-wave interaction is smaller contribution than other channels.
Thus, $D^*\pi$ and $D_2^*(2460)\pi$ could be two key decay channels when one experimentally searches for these two $2^-$ charmed mesons. If taking the typical value $m=2.76$ GeV for these two $2^-$ charmed mesons, the total decay widths can reach up to $103$ MeV and $143$ MeV for $2^-(1``^3D_2")$ and $2^-(1``^1D_2")$ states, respectively.

\begin{center}
\begin{figure}[htb]
\begin{tabular}{cccc}
\scalebox{0.65}{\includegraphics{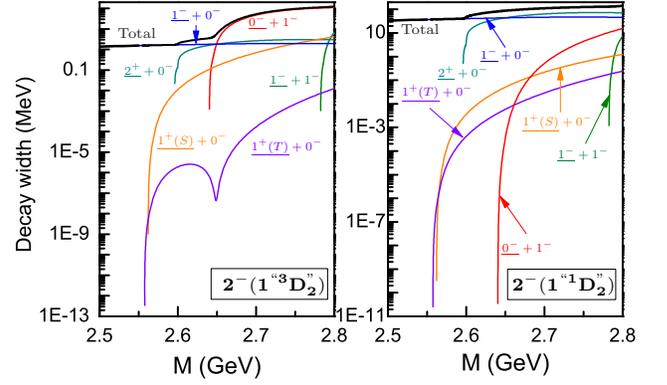}}
\end{tabular}
\caption{(Color online.) The figure shows the mass dependence of the total two-body and partial strong decay of two $1D$ charmed mesons with $J^P=2^-$, which are mixing of $1^1D_2$ and $1^3D_2$ states with definitions
$|``^3D_2"\rangle=\sqrt{3/5}|^3D_2\rangle-\sqrt{2/5}|^1D_2\rangle$ and
$|``^1D_2"\rangle=\sqrt{2/5}|^2D_2\rangle+\sqrt{3/5}|^1D_2\rangle$ \cite{Close:2005se,Matsuki}. Here, the mass range of these two $2^-$ charmed mesons is taken as $2.5 - 2.8$ GeV, and the $J^P$ quantum numbers of charmed mesons of final state are marked by underline. \label{NR-3} }
\end{figure}
\end{center}

In summary, three newly observed charmed resonances, $D(2550)$, $D(2610)$, and $D(2760)$, have been for the first time assigned as $2S$ and $1D$ charmed meson by the analysis of mass spectrum as well as the calculation of two-body strong decay. Concretely, $D(2610)$ should be the mixing of $D(2^3S_1)$ and $D(1^3D_1)$ charmed mesons with spin-parity $J^P=1^-$, where pure $1^-(2^3S_1)$ state assignment to $D(2610)$ has been fully excluded in this letter. There exits two structure assignments to $D(2760)$, i.e., the orthogonal cousin of $D(2610)$ or $3^-(1^3D_3)$, which can be distinguished by further study on the main decay channels of $D(2760)$ in future experiment.
{Although $D(2550)$ is seemingly explained as $2^1S_0$ state under the analysis of the mass spectra, we have found discrepancy between theory and the experiment on the width for $D(2550)$. Our theoretical calculation of its decay width is far less than the observed one by BaBar. This may be due to our quark model assumption for $D(2550)$ or due to the assignment $2^1S_0$ to $D(2550)$.}
Moreover, the decay behavior of the remaining two $1D$ charmed mesons have also been predicted, which provides valuable information for future experiments to search for all $1D$ charmed states.

Just because of the similarity between charmed and charmed-strange mesons, this study will shed light on the underlying properties of the observed charmed-strange states $D_{s1}(2710)$ and $D_{sJ}(2860)$ \cite{:2009di} due to two facts which may reflect the global flavor $SU(3)$ recovery \cite{Matsuki:2008ms} : the mass gap between $D(2760)$ and $D(2610)$ surprisingly agrees with that of $D_{sJ}(2860)$ and $D_{s1}(2710)$, both of which are about 150 MeV; 
$D_{sJ}(2860)/D_{s1}(2710)$ have almost the same mixing angle between $1^3S_1$ and $1^3D_1$ states as that of charmed cousins $D(2760)/D(2610)$ if considering the mixing of two $1^-$ states to explain $D(2760)/D(2610)$ and $D_{sJ}(2860)/D_{s1}(2710)$ \cite{Li:2009qu}.
More abundant experimental phenomena together with further efforts on phenomenological study will contribute to our understanding of heavy-light meson system.

\section*{Acknowledgements}

This project is supported by the National Natural
Science Foundation of China under Grants No. 10705001, No. 11035006 and the Ministry of Education of China (FANEDD under Grant No. 200924, DPFIHE under Grant No. 20090211120029, NCET under Grant No. NCET-10-0442, the Fundamental Research Funds for the Central Universities under Grant No. lzujbky-2010-69). One of the authors (TM) would like to express his sincere thanks to Prof. Xiang Liu who provided nice atmosphere while he stayed at Lanzhou University.

\vfil

\end{document}